\documentclass[conference]{IEEEtran}
\usepackage{algorithmic}
\usepackage{amsmath,amssymb,amsfonts}
\usepackage{booktabs}
\usepackage{cite}
\usepackage{graphicx}
\usepackage{soul}
\usepackage{textcomp}
\usepackage{xcolor}
\def\BibTeX{{\rm B\kern-.05em{\sc i\kern-.025em b}\kern-.08em
    T\kern-.1667em\lower.7ex\hbox{E}\kern-.125emX}}
\newcommand{\acf}{r_{xx}[k]}

\newcommand{\norm}{\nu}
\newcommand{\periodogram}{\hat{P}(u)}

\newcommand{\psd}{P_{xx}(u)}

\begin{document}
\setlength{\abovedisplayskip}{7.5pt}
\setlength{\belowdisplayskip}{7.5pt}
\title{Product Processing for Tapered Sparse Arrays}
\author{\IEEEauthorblockN{Daniel Sartori, Kaushallya Adhikari}
	\IEEEauthorblockA{\textit{Electrical, Computer, and Biomedical Engineering Department}\\ {University of Rhode Island, Kingston, RI 02881 USA} \\
		\{dsartori,kadhikari\}@uri.edu}
}
\maketitle
\begin{abstract}
The product processor output has recently been introduced as a spatial power spectral density estimate, unifying product arrays such as coprime arrays, nested arrays, and standard uniform line arrays. The expected value and covariance function of this estimate for a white Gaussian process was derived in previous work over these various array configurations. However, this prior work used a uniform taper in all cases. In this paper, we show that when product arrays are windowed with non-uniform tapers, the expected value of the product processor output is the convolution of the true spatial power spectral density with the spatial Fourier transform of the difference coarray. This expected value makes a Fourier transform pair with a spatial autocorrelation estimate obtained by windowing the true autocorrelation function. We also derive the covariance function of the product processor output with non-uniform tapers, and compare these derived statistics for the aforementioned array geometries. Also, in prior work, the moments were provided only for linear arrays; this paper extends the estimation results to multidimensional arrays.
\end{abstract}
\section{Introduction}
\label{sec:intro}
Sensor arrays are important in applications such as sonar, radar, communications, and seismology. These arrays spatially sample their environment allowing for direction of arrival (DoA) estimation of impinging signals in ambient noise. The DoA estimates suffer from the \textit{aliasing} artifact if the intersensor spacing is larger than half the wavelength ($\lambda/2$) of the propagating signal \cite{VanTrees,JD,HoctorKassam}. In contrast, increasing this spacing lengthens the overall array aperture thereby improving the resolution. In a standard uniform line array (ULA), the sensor positions lie on a single axis, and intersensor spacing conforms to the half-wavelength constraint. However, we can create cost effective arrays with an average intersensor spacing larger than $\lambda/2$ using the same aperture length. Such arrays are called sparse arrays.\par
Product processing is an effective strategy for resolving the ambiguities created when the array spacing exceeds the standard. It involves populating a uniform aperture using two subarrays and multiplying their outputs to obtain a \textit{product spectrum} \cite{berman}. Recent examples of product processing on sparse arrays include nested and coprime arrays, which are formed by interleaving two ULAs~\cite{nested1,nested2,VandP1,VandP2}. This paper focuses on these two sparse arrays.\par
In prior work, it was proven that estimating the spatial cross-correlation function between two subarrays is equivalent to multiplying their conventional beamforming (CBF) outputs to obtain the product processor output (PPO). This is equivalent to the spatial periodogram in~\cite{psdestimation}. \cite{psdestimation} also derives the expected value of the PPO for any propagating signal and the covariance of spatially white Gaussian noise, but the authors only incorporate uniform windows in each case. Non-uniform windows are essential to prevent the masking of weak signals due to the sidelobes of strong interferers \cite{Kay}. 
In addition, the results in \cite{psdestimation} do not apply to multidimensional arrays.\par
\begin{figure}[t]
	\centering
	\includegraphics[scale=0.21,trim = 0cm 0cm 0cm 0cm]{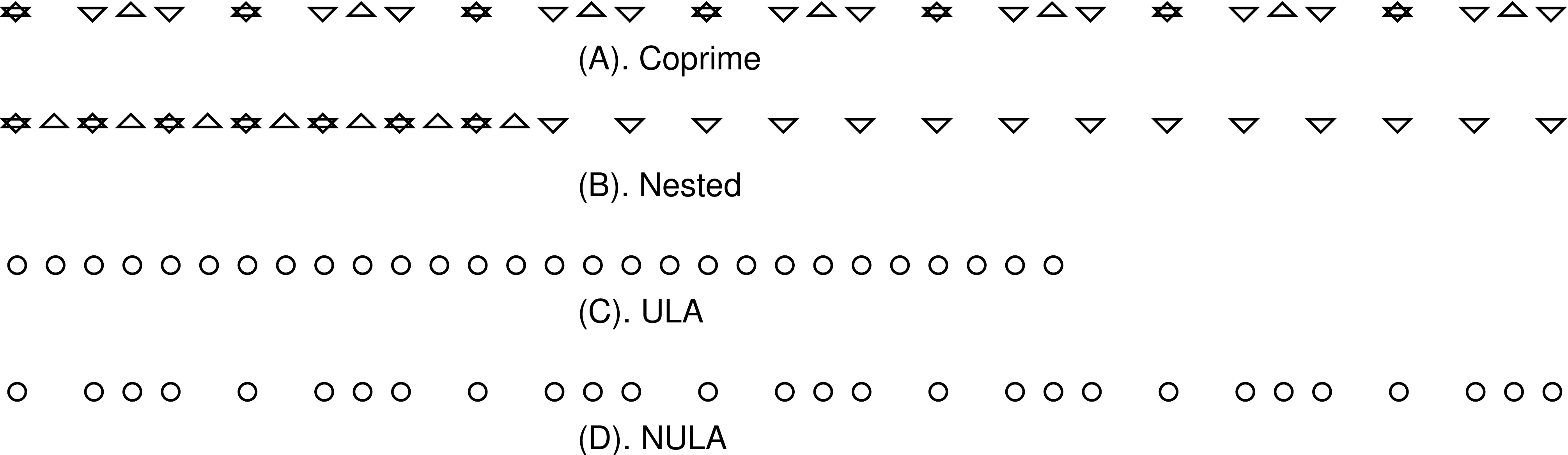}
	\caption{(A). A coprime array with $28$ sensors. Subarray~A has $14$ sensors with USF equal to $3$, and Subarray~B has $21$ sensors with USF equal to $2$. (B). A nested array with 28 sensors. Subarray~A has $14$ sensors with USF equal to $1$, and Subarray~B has $21$ sensors with USF equal to $2$. (C). A standard ULA with $28$ sensors. (D). A NULA with 28 sensors. \label{geometries}}
\end{figure}
Product processing and min processing are the predominant CBF-based methods for coprime and nested arrays \cite{VandP1,nested1,KBW,shadings,MartinoIodice,adhikariJasa1,chavali2,
icasspdetection,liubuck4,detection2019}. Various aspects of these two processors are compared in depth in \cite{chavali2,liubuck4,WageMagazine}, with product processing having the advantage of being less vulnerable to \textit{crossterms} than min processing \cite{chavali2,liubuck4,AdhikariAccess1}. Other methods that have been popular for coprime and nested arrays are subspace-based algorithms such as MUSIC and ESPRIT \cite{VandP1,nested1,AccessCoprime1,asaconf1,AccessCoprime2,asaconf2,AccessCoprime6,AccessCoprime8,AccessCoprime11,AccessCoprime17,AdhikariNAECON}. For lattice-imposed planar arrays, these methods exhibit very low resolution when compared with a PPO due a narrow filled coarray \cite{AdhikariAccess2}. Because these algorithms process an estimate of the signal correlation matrix to obtain an approximate DoA, the statistical properties of the PPO also provide a means to analyze subspace-based algorithms.\par
Our specific contributions in this paper are
\begin{enumerate}
\item Derive the expected value of a PPO using general tapers.
\item Derive the covariance of this PPO for spatially white Gaussian signals for all tapers.
\item Extend the PPO results to multidimensional arrays and find the expected value.
\end{enumerate}
\textit{Conventions:} $^T$ denotes Transpose; $^H$ denotes Hermitian; $\mathcal{F}\lbrace\hspace{0.25em}\rbrace$ denotes Fourier Transform operator; bold-faced letters represent vectors; if $\mathbf{z}$ is a vector, then $z[l]$ represents the $l^{th}$ element of vector $\mathbf{z}$; $\mathbf{a}\odot\mathbf{b}$ denotes the Hadamard product (element-wise product) of $\mathbf{a}$ and $\mathbf{b}$; vectors $\mathbf{m}$, $\mathbf{n}$, $\mathbf{u}$, $\mathbf{k}$, and $\mathbf{l}$ without subscript are Cartesian coordinate vectors in $\mathbb{R}^3$,for example $w[\mathbf{m}] = w[m_x,m_y,m_z]$.
\color{black}
\vspace*{-2mm}
\section{Product Arrays}
\label{sec:prod}
A product array comprises two linear subarrays, hereafter referred to as Subarray~A and Subarray~B, where the individual CBF outputs are multiplied to obtain the output. Each subarray could be a standard ULA, a sparse ULA, or a sparse non-uniform line array (NULA).\par
Coprime arrays are an example of a product array where both subarrays are sparse colinear ULAs~\cite{VandP1}, and the undersampling factors (USF) in the two subarrays must be coprime \cite{VandP1,VandP2,shadings,KBW}. Fig.~\ref{geometries}A depicts a coprime array where Subarray~A and Subarray~B have $14$ and $21$ sensors respectively, with USFs equal to $3$ and $2$ respectively.\par
Another popular product array is a nested array, in which Subarray~A is a standard ULA, and Subarray~B is a sparse ULA~\cite{nested1,nested2}. Fig.~\ref{geometries}B depicts a nested array in which Subarray~A has $14$ sensors with USF equal to $1$, and Subarray~B has $21$ sensors with USF equal to $2$.\par
Fig.~\ref{geometries}C depicts a product array interpretation of a standard ULA as seen in~\cite{psdestimation} where the $28$ sensor locations are all shared sensors.\par
Similarly, Fig.~\ref{geometries}D depicts a NULA where the two subarrays have identical geometry, and all sensors are shared. Note that the sensor locations of the NULA in Fig.~\ref{geometries}D are the same as the sensor locations of the coprime array in Fig.~\ref{geometries}A.
\section{Signal Model and PPO}
\label{sec:generalization}
We assume that the array is along the positive $z$-axis with the first sensor at $z=0$. When a plane-wave signal impinges on the array at an angle of $\theta_s$ with respect to the array axis, the signal vector received by Subarray~A is given by
\begin{equation}
\label{data1}
\mathbf{x}_A=(s\mathbf{v}_{A}(\theta_s)+\mathbf{n}_A)\odot\boldsymbol{\kappa}_A,
\end{equation}
where $s$ is a complex random variable representing signal amplitude, $\mathbf{n}_A$ is a complex random vector of noise values at $M_e$ possible sensor locations, and $\boldsymbol{\kappa}_A$ is an $M_e$-element indicator vector used to index the sensor positions in multiples of $\lambda/2$. $\boldsymbol{\kappa}_A$ contains a one where sensors are present and a zero where they are skipped. $\mathbf{v}_{A}(\theta_s)$ is the array manifold vector for Subarray~A, given by
\begin{equation}
\label{data1a}
\mathbf{v}_{A}(\theta_s)=[e^{j\pi u_s 0}\hspace*{0.5em}e^{j\pi u_s 1}\hspace*{0.5em}e^{j\pi u_s 2}\hdots e^{j\pi u_s (M_e-1)}]^T
\end{equation}
where $u_s=\cos(\theta_s)$ is the direction cosine, and $M_e-1$ is the aperture of Subarray~A in multiples of $\lambda/2$. The direction vector $\mathbf{v}_{A}(\theta_s)$ and the noise vector $\mathbf{n}_A$ both have $M_e$ elements. However, the number of sensors in Subarray~A could be anywhere between $1$ and $M_e$, where the sparsity is determined by $\boldsymbol{\kappa}_A$. If the number of sensors is $M_e$, then Subarray~A is a full ULA, and $\boldsymbol{\kappa}_A$ is simply a vector of ones. Similarly, the signal vector received by Subarray~B is given by
\begin{equation}
\label{data2}
\mathbf{x}_B=(s\mathbf{v}_{2}(\theta_s)+\mathbf{n}_B)\odot\boldsymbol{\kappa}_B
\end{equation}
where $\mathbf{v}_{B}(\theta_s)$ is the array manifold for direction $\theta_s$, $\mathbf{n}_B$ is the noise vector, and $\boldsymbol{\kappa}_B$ is the indicator vector for Subarray~B. These vectors are $N_e$-element vectors, since the aperture of Subarray~B is $N_e-1$ multiples of $\lambda/2$.\par

The product processor first applies tapers $\mathbf{w}_1$ and $\mathbf{w}_2$ to the signals $\mathbf{x}_A$ and $\mathbf{x}_B$, where $\mathbf{w}_1$ and $\mathbf{w}_2$ are $M_e$ and $N_e$ in length, respectively. If sensors are missing, the corresponding elements of $\mathbf{w}_1$ and $\mathbf{w}_2$ are zero. At other locations, the elements of $\mathbf{w}_1$ and $\mathbf{w}_2$ depend on the taper. To compute the output corresponding to the steered direction, $\theta$, the product processor also applies vectors $\mathbf{v}_{1,\theta}$ and $\mathbf{v}_{2,\theta}$ with elements $v_{1,\theta}[m]=e^{j\pi u m}$ and $v_{2,\theta}[n]=e^{j\pi u n}$ to generate the subarray outputs given by
\begin{equation}
\label{subarrayoutputs}
\begin{split}
&y_1=\sum_{m=0}^{M_e-1} w_1[m]v_{1,\theta}[m]^*x_1[m]\text{ and}\\
&y_2=\sum_{n=0}^{N_e-1}w_2[n]v_{2,\theta}[n]^*x_2[n],
\end{split}
\end{equation}
The PPO, $\periodogram=y_1y_2^*/\norm$, has the structure of a periodogram given by
\begin{equation}
\label{spatialperiodo}
\periodogram = \dfrac{1}{\norm}\sum\limits_{m=0}^{M_e-1}\sum\limits_{n=0}^{N_e-1}w_1[m]w_2[n]^* x_1[m]x_2^*[n] e^{-j\pi u(m-n)}
\end{equation}
where the normalization constant, $\norm$, is given by
\begin{equation}
\label{nu}
\norm=\sum_{k=1}^{M_e}w_1[k]w_2^*[k]
\end{equation}
This choice of the normalization constant makes the PPO unbiased for spatially white noise. An important property of~\eqref{nu} is that changing the upper limit to $N_e$ would not change the value of $\norm$. The reason for this is that when $M_e\geq N_e$, the value of $w_2^*[k]$ is $0$ for all $k>N_e$, and therefore the product is $0$ for all $k>N_e$. Similarly, when $M_e\leq N_e$, $w_1[k]=0$ for all $k>M_e$, and so the product becomes $0$ for all $k>M_e$.\par
The PPO in~\eqref{spatialperiodo} does not make any assumptions regarding the geometries of the subarrays or their tapers. This is a generalized form of the results obtained in~\cite{psdestimation}, which only includes subarrays that are ULAs with uniform tapers.
\section{Mean and Covariance of PPO}
\label{sec:properties}
\subsection{Spatial Autocorrelation Function Estimate}
\label{sec:acf}
The Wiener-Khinchin theorem in temporal spectral estimation states that the power spectral density (PSD) of a wide sense stationary (WSS) signal and its autocorrelation function (ACF), $r_{xx}[k]$, are related by the Fourier transform~\cite{Kay}. The PPO makes the Fourier transform pair with a spatial ACF estimate, $\hat{r}[k]$. With $k=m-n$, the expression for $\periodogram$ in \eqref{spatialperiodo} can be rewritten as $\periodogram $
\begin{equation}
\label{pa}
\begin{split}
= &\sum_{k=-(N_e-1)}^{M_e-1}\sum_{l}(w_1[l]x_1[l])(w_2[l-k]x_2[l-k])^*\dfrac{e^{-j\pi uk}}{\norm} \\
= &\sum_{k=-(N_e-1)}^{M_e-1}\dfrac{(w_1[k]x_1[k])\star (w_2[-k]x_2[-k])^*}{\norm}e^{-j\pi uk},
\end{split}
\end{equation}
where $\star$ denotes linear convolution. The implicit ACF estimate, $\hat{r}[k]$, corresponding to $\periodogram$ is given by
\begin{equation}
\label{pa2}
\begin{split}
&\hat{r}[k]=(w_1[k]x_1[k])\star (w_2[-k]x_2[-k])^*/\norm\hspace{2mm}\text{and}\\ &\hat{P}(u)=\sum_{k=-(N_e-1)}^{M_e-1}\hat{r}[k]\exp(-j\pi u k).
\end{split}
\end{equation}
The expected value of the ACF estimate is
\begin{equation}
\label{meanacf}
\begin{split}
&E\lbrace \hat{r}[k]\rbrace=  r_{xx}[k]\dfrac{1}{\norm}\displaystyle\sum_{l}w_1[l]w_2^*[l-k]=r_{xx}[k]w_c[k],\\
\end{split}
\end{equation}
where $r_{xx}[k]$ is the true spatial ACF and $w_c[k]$ is the weighting function made from the normalized convolution of the two tapers. The weighting function reverses and conjugates the tapers such that $w_c[k]=w_1[k]\star w_2^*[-k]/\norm$. Since the expected value of the ACF estimate is a weighted ACF rather than the true one, the ACF estimate is generally biased. When both subarrays have uniform tapers, the weighting function, $w_c[k]$, becomes the difference coarray of the two subarrays \cite{psdestimation}. The value of $w_1[k]\star w_2^*[-k]$ at $k=0$ is $\sum_{l=1}^{M_e}w_1[l]w_2^*[l]$. Hence, normalizing the convolution by $\norm$ forces $w_c[0]$ to be equal to $1$. Since the true ACF of a spatially white noise process is of the form $r_{xx}[k]=\sigma^2\delta (k)$, where $\sigma^2$ is the white noise variance, the expected value of the ACF estimate for the white noise is $E\lbrace \hat{r}[k]\rbrace=\sigma^2\delta(k)w_c[0]=\sigma^2\delta(k)=r_{xx}[k].$ Thus, the ACF estimate is unbiased for a spatially white noise process for any subarray tapers.
\subsection{Expected Value of the PPO}
\label{sec:expected}
Intuitively, the PPO is a biased estimate of the true spatial PSD, since its inverse Fourier transform, the ACF estimate, is biased. Taking the expected value of the PPO in \eqref{pa2} and substituting the expected value of the ACF estimate from \eqref{meanacf}, the expected value of the PPO becomes
\begin{equation}
\label{meanexpression}
\begin{split}
E\left\lbrace\periodogram \right\rbrace &= \sum_{k=-(N_e-1)}^{M_e-1}w_c[k]\acf e^{-j\pi u k}\\
&=\mathcal{F}\lbrace \acf w_c[k]\rbrace=\psd\circledast W_c(u),
\end{split}
\end{equation}
where $\circledast$ denotes periodic convolution. Hence, the expected value is the smoothed version of the true PSD and biased with respect to the Fourier transform of the weighting function, $w_c[k]$.\par
For spatially white noise, the area under $W_c(u)$ is $\textstyle\frac{1}{2}\int_{-1}^1 W_c(v)du=w_c[0]=1$. Since the true PSD, $P_{xx}[u]$, is a constant, the expected value equals the true value.  and the PPO becomes an unbiased estimate for any taper, which is consistent with Section~\ref{sec:acf}.
\subsection{Covariance of the PPO}
\label{sec:variance}
For spatially white Gaussian noise, the covariance between the  PPO values at $u=u_1$ and $u=u_2$ is $C(\Delta u =u_1-u_2)$
\begin{equation}
\label{covariance}
\begin{split}
&\frac{\sigma^4}{|\norm|^2} W_1(\Delta u)\circledast W_1^*(-\Delta u)( W_2(\Delta u)\circledast W_2^*(-\Delta u))^*,
\end{split}
\end{equation}
where $W_1(\Delta u)$ and $W_2(\Delta u)$ are the Fourier transforms of the subarray tapers $w_1[m]$ and $w_2[n]$ and $\sigma^2$ is the white noise variance. The complete derivation is given in Appendix~\ref{appendix:gen}. The variance of the PPO, as obtained by substituting $\Delta u=0$  in~\eqref{covariance} and simplifying, is
\begin{equation}
C(0)=\dfrac{\sigma^4}{|\norm|^2}\Biggl\lbrace\sum_{k=0}^{M_e-1}|w_1[k]|^2\Biggr\rbrace \Biggl\lbrace\sum_{l=0}^{N_e-1}|w_2[l]|^2\Biggr\rbrace ^*.
\end{equation}
In an array where Subarray~A and Subarray~B are equal, as in Fig.~\ref{geometries}C and Fig.~\ref{geometries}D, both the terms in braces are equal to $\norm$. Therefore, for white noise with variance $\sigma^2$, the PPO variance is $\sigma^4$, which is the square of the expected value. For an array where Subarray~A and Subarray~B are not equal, the variance can be higher than $\sigma^4$. Hence, the variance of the PPO is high, even for the conventional geometry of Fig.~\ref{geometries}C, and does not necessarily decrease with an increase in the number of sensors and array aperture.\par
To improve the variance of the spatial PSD estimate, we compute the PPO for each one of $K$ independent snapshots of the propagating signal, and then evaluate the average. The expected value of the PPO does not change with averaging, and is still given by the periodic convolution of the true PSD with the Fourier transform of the weighting function. However, averaging will reduce the variance of the PPO by a factor of $K$.
%
\subsection{Comparison of PPO for Different Geometries}
\label{sec:eg}
\begin{figure}[t]
	\centering
	\includegraphics[scale=0.42,trim = 0cm 19cm 0cm 2.2cm]{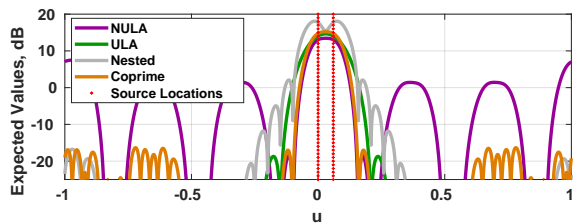}	\caption{Comparison of the expected values of the PPO for a NULA (purple), ULA (green), nested array (gray), and a coprime array (brown), all using Hann windows. The nested array is able to resolve the two source directions, but other arrays are not. \label{meanHann}}
\end{figure}

Next, we compare the statistics of four examples of PPOs for the geometries depicted in Fig.~\ref{geometries}. The total number of sensors for each array has been restricted to $28$ in order to facilitate a fair comparison. The sparse arrays have been designed with equal aperture, and hence, equal resolution. Fig.~\ref{meanHann} compares the expected values of the PPO for the product arrays using Hann tapering. The coprime array, nested array, and NULA have equal null-to-null main lobe width (MLW) due to their equal apertures. The ULA has a wider null-to-null main lobe width because its aperture is the shortest. However, the monotonically decreasing side lobe behavior of the ULA looks better than the other geometries.

\section{PPOs for Spatially Colored Processes}
\label{sec:colored}
The closed form expressions for the expected value and variance of the PPO derived in Section~\ref{sec:properties} facilitate the comparison of various popular geometries in ways that were not previously possible. In temporal spectral estimation, where the windows are uniform, the expected value of the PPO is the convolution of the true PSD with the Fourier transform of a Bartlett window \cite{Kay}. When the main lobe width (MLW) of the Bartlett window is significantly narrower than the narrowest peak in the true PSD, the PPO is approximately unbiased even for a colored process~\cite{JenkinsWatts, Kay}.

A parallel discussion can be established for the spatial case. Since the expected value of the generalized spatial PPO is the convolution of the true PSD with the function $W_c(u)$, the effect of the function $W_c(u)$ is to smear the average PPO. Thus, when the MLW of $W_c(u)$ is much narrower than the peaks in the true PSD of a spatially colored process, the bias in the estimate is expected to be insignificant. 

\begin{figure}[t]
	\centering
	\includegraphics[scale=0.4,trim = 0cm 17cm 0cm 2cm]{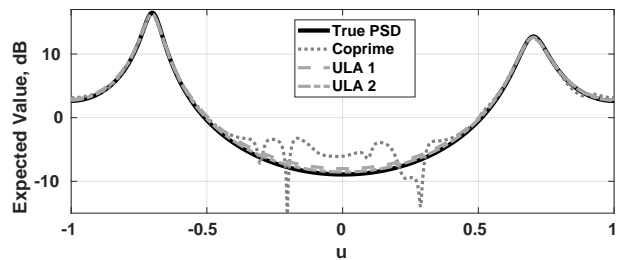}
	\caption{Comparison of the expected values of the PPO with the true PSD (solid black) for a coprime array (gray dotted), ULA (gray dash-dot) that matches the number of sensors of the coprime array, and ULA (gray dashed) that matches the resolution of the coprime array, using uniform windows.  \label{coloredfig}}
\end{figure}

This notion is analyzed for two different array designs in Fig.~\ref{coloredfig}. The solid black line depicts the true PSD of a spatially colored process. At direction cosines $u=-0.7$ and $u=0.7$, the $3$~\rm{dB} bandwidths of the peaks  are $0.04$ and $0.067$, respectively. Consider a coprime array where Subarray~A has $40$ sensors with USF equal to $5$ and Subarray~B has $50$ sensors with USF equal to $4$. The coprime array has a total of $80$ sensors, a MLW of $0.02$, and a PSL height of $-13.3$ \rm{dB}. The MLW of the coprime array is narrower than the narrowest peak in the true PSD. Hence, the bias in the coprime array's PPO estimate is expected to be low. This low bias is clear in Fig.~\ref{coloredfig}, since the two peaks in the true PSD are well matched by the coprime array, shown by the dotted line.
 
The dashed ULA~1 line in Fig.~\ref{coloredfig} illustrates the expected value of the ULA PPO where the number of sensors in the ULA and coprime array both $80$. Hence, the null-to-null MLW of the ULA is $0.05$, which, though wider than that of the coprime array, is comparable to the peaks in the true PSD. The ULA is also able to detect the two peaks in the true PSD very well. However, the overall bias for the coprime array seems higher than the overall bias for the ULA.\par
If we create a ULA with the same resolution (same MLW) as the coprime array, it would require $200$ sensors. For this ULA, the bias in the PPO estimate, as shown by the dash-dot ULA~2 line in Fig.~\ref{coloredfig}, is negligible. Thus, among PPOs with identical subarrays,  when $W_c$ has a MLW that is narrower than the peaks in the true PSD, the PPO is almost unbiased. However, for a design where the subarrays are different, like a coprime array, this notion is only partially true. The bias in the actual peaks of the estimate seems negligible, but the bias for other direction cosines (the region between $u=-0.45$ and $u=0.45$) is substantially high. Even with very narrow MLWs of $W_c(u)$, some designs such as the coprime array fail to produce approximately unbiased PPO estimates; this can be explained by the side lobe pattern of $W_c(u)$.\par
\begin{figure}[t]
	\centering
	\includegraphics[scale=0.43,trim = 0cm 15cm 0cm 1cm]{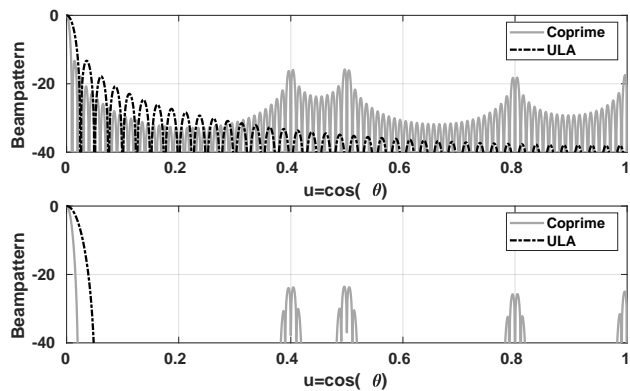}
	\caption{Comparison of the beampatterns of a coprime array (solid gray) and a ULA (dashed-dot black). Top Panel: All subarrays use uniform windows. Bottom Panel: All subarrays use hamming windows. \label{beampatternscsaula}}
\end{figure}

Fig.~\ref{beampatternscsaula} depicts the function $W_c(u)$ of the coprime array and the full ULA with $80$ sensors (from Fig.~\ref{coloredfig}). 
The MLW of the coprime array is slightly narrower than the ULA, and the PSL heights of the two arrays are almost equal. However, the coprime array exhibits poor side lobe behavior overall. The ULA side lobe peaks decrease monotonically, whereas the coprime array does not exhibit this behavior. When convolving $W_c(u)$ with the true PSD, $P_{xx}(u)$, the MLW of $W_c(u)$ does not smear the true PSD that much, but the smoothing caused by the side lobes can cause substantial bias in the PPO estimate. However, when the subarrays of the coprime array are windowed by a non-uniform window (e.g. hamming), the overall side lobe behavior of the coprime array is more comparable with that of the ULA with the same window, as depicted by the bottom panel of Fig.~\ref{beampatternscsaula}.\par

\begin{figure}[t]
	\centering
	\includegraphics[scale=0.4,trim = 1cm 20cm 0cm 1.6cm]{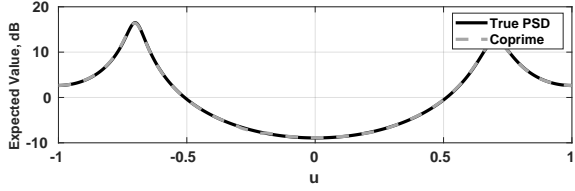}
	\caption{Comparison of the expected values of the PPO with the true PSD (solid black) for a coprime array (gray dash-dot), where all subarrays use hamming windows. \label{coloredfig2}}
\end{figure}

Subsequently, the convolution of the resulting $W_c(u)$ with the true PSD does not cause substantial smoothing, and the bias is negligible, as shown in Fig.~\ref{coloredfig2}. This proves that for an array design where the two subarrays are not equal, as a coprime array and a nested array, if the MLW of $W_c(u)$ is narrower than the narrowest peak in the true PSD, and the total side lobe area is low, then the PPO estimate will be approximately unbiased, for a spatially colored process.

\section{PPO for Multidimensional Arrays}
\label{sec:multidim}
The PPO definition and properties extend neatly to multidimensional arrays. The CBF beampatterns of two three-dimensional subarrays are 
\begin{equation}
\nonumber
\begin{split}
y_1=&\sum_{m_x=0}^{M_{x}-1}\sum_{m_y=0}^{M_y-1}\sum_{m_z=0}^{M_z-1} w_1[\mathbf{m}]v_{1,\theta}[\mathbf{m}]^* x_1[\mathbf{m}] \text{ and}
\end{split}
\end{equation}

\begin{equation}
\nonumber
\begin{split}
y_2=&\sum_{n_x=0}^{N_x-1}\sum_{n_y=0}^{N_y-1}\sum_{n_z=0}^{N_z-1}w_2[\mathbf{n}]v_{2,\theta}[\mathbf{n}]^* x_2[\mathbf{n}],
\end{split}
\end{equation}
where $v_{1,\theta}[\mathbf{m}]=e^{j\pi(\mathbf{u}^T\mathbf{m})}$ and $v_{2,\theta}[\mathbf{n}]=e^{j\pi(\mathbf{u}^T\mathbf{n})}$. The quantities $u_x=\sin(\theta)\cos(\phi),$ $u_y=\sin(\theta)\sin(\phi)$, and $u_z=\cos(\theta)$ are the direction cosines with respect to the $x$, $y$, and $z$ axes. The PPO is now a multivariate function given by
\begin{equation}
\nonumber
\begin{split}
&\hat{P}(\mathbf{u}) = \dfrac{1}{\gamma}\sum_{m_x=0}^{M_{x}-1}\sum_{m_y=0}^{M_y-1}\sum_{m_z=0}^{M_z-1}\sum_{n_x=0}^{N_x-1}\sum_{n_y=0}^{N_y-1}\sum_{n_z=0}^{N_z-1}\times \\
&\Biggl\lbrace w_1[\mathbf{m}]w_2^*[\mathbf{n}]x_1[\mathbf{m}]x_2^*[\mathbf{n}]   e^{-j\pi (\mathbf{u}^T\mathbf{(m-n)})}\Biggr\rbrace,
\end{split}
\end{equation}
where $\gamma=\sum_{c_x=1}^{M_x}\sum_{c_y=1}^{M_y}\sum_{c_z=1}^{M_z}w_1[\mathbf{c}]w_2^*[\mathbf{c}]$ is the multidimensional extension of the normalization factor $\norm$. The PPO above simplifies to  the Fourier transform of the function $(w_1[\mathbf{k}]x_1[\mathbf{k}])\star (w_2[-\mathbf{k}]x_2[-\mathbf{k}])^*/\gamma$, which is the implicit ACF estimate $\hat{r}[\mathbf{k}]$. The full derivation is given in Appendix \ref{sec:multiproof}.

The expected value of the multivariate ACF estimate is
\begin{equation}
\label{multimeanacf}
E\lbrace \hat{r}[\mathbf{k}]\rbrace=r_{xx}[\mathbf{k}]w_c[\mathbf{k}].
\end{equation}
See Appendix \ref{sec:multiproof} for the proof. The multivariate weighting function in \eqref{multimeanacf} is
\begin{equation}
\label{multiwc}
w_c[\mathbf{k}]=w_1[\mathbf{k}]\star w_2^*[-\mathbf{k}]/\gamma.
\end{equation}
As with linear arrays, the expected value of the multivariate ACF estimate is not equal to the true autocorrelation function. Hence, the ACF and the PPO are biased estimates.

\section{Conclusion}
In this paper, we derived the first two moments of the PPO when the subarrays are tapered with non-uniform windows. We evaluated and compared the PPO statistics of four different arrays. We discussed the conditions that produce unbiased PPO estimates. The results were also extended to multidimensional arrays.
\appendices

\section{Moments of the PPO}

\subsection{Covariance of the PPO}
\label{appendix:gen}

The second moment of the PPO is $E\left\lbrace \hat{P}(u_1)\cdot\hat{P}^*(u_2)\right\rbrace$
\begin{align}
\nonumber
\begin{split}
&=\sum_{k=0}^{M_e-1}\sum_{l=0}^{N_e-1} \sum_{m=0}^{M_e-1}\sum_{n=0}^{N_e-1}\Biggl\lbrace\frac{1}{\norm\norm^*}w_1[k]w_2^*[l]w_1^*[m]w_2[n]\\
&\times E\left\lbrace x[k]x^*[l]x^*[m]x[n]\right\rbrace \times e^{j\pi(u_1(l-k)+u_2(m-n))}\Biggr\rbrace
\end{split}
\end{align}
For a white Gaussian noise with power $\sigma^2$, $E\lbrace \hat{P}(u_1)\cdot\hat{P}^*(u_2)\rbrace$
\begin{align}
\nonumber
\begin{split}
&=\dfrac{\sigma^4}{|\norm|^2}\sum_{k=0}^{M_e-1}\sum_{l=0}^{N_e-1} \sum_{m=0}^{M_e-1}\sum_{n=0}^{N_e-1}w_1[k]w_2^*[l]w_1^*[m]w_2[n]\times\\
&\Biggl(\delta[k-l]\delta[n-m]+\delta[k-m]\delta[n-l]\Biggr)e^{j\pi(u_1(l-k)+u_2(m-n))}
\end{split}
\end{align}
Splitting $E\left\lbrace \hat{P}(u_1)\cdot\hat{P}^*(u_2)\right\rbrace$ into 2 terms, where $A$ is
\begin{equation}
\nonumber
\begin{split}
&A=\frac{\sigma^4}{|\norm|^2}\sum_{k=0}^{M_e-1}\sum_{l=0}^{N_e-1} \sum_{m=0}^{M_e-1}\sum_{n=0}^{N_e-1}w_1[k]w_2^*[l]w_1^*[m]w_2[n]\\
&\Biggl\lbrace\delta[k-l]\delta[n-m])\times e^{(j\pi(u_1(l-k)+u_2(m-n))}\Biggr\rbrace
\end{split}
\end{equation}
and $B$ is
\begin{equation}
\nonumber
\begin{split}
&B=\frac{\sigma^4}{|\norm|^2}\sum_{k=0}^{M_e-1}\sum_{l=0}^{N_e-1} \sum_{m=0}^{M_e-1}\sum_{n=0}^{N_e-1}w_1[k]w_2^*[l]w_1^*[m]w_2[n]\\
&\Biggl\lbrace\delta[k-m]\delta[n-l]\times e^{j\pi(u_1(l-k)+u_2(m-n))}\Biggr\rbrace,\\
\end{split}
\end{equation}
the second moment simplifies to $=A+B.$

The term $\delta[k-l]\delta[n-m]$ in $A$ is non-zero ($1$, specifically) only when $k=l$ and $n=m$ at the same time. This condition, $k=l$ and $n=m$, is true at each shared sensor location. Also, when $k=l$ and $n=m$, the term $e^{(j\pi(u_1(l-k)+u_2(m-n))}$ is $1$. Hence, the quantity $A$ can be simplified to
\begin{equation}
\nonumber
A=\frac{\sigma^4}{|\norm|^2}\biggl(\sum_{k=0}^{M_e-1}w_1[k]w_2^*[k]\biggr)\biggl(\sum_{m=0}^{M_e-1}w_1[m]w_2^*[m]\biggr)^*.
\end{equation}
Noting that each term in parentheses in the above equation is $\norm$, the expression for $A$ simplifies to $A=\sigma^4.$

For $B$, rearranging and replacing $u_1-u_2$ with $\Delta u,$ $B=$ 
\begin{equation}
\nonumber
\begin{split}
\dfrac{\sigma^4}{|\norm|^2}&\sum_{k=0}^{M_e-1}w_1[k]w_1^*[k]e^{-j\pi\Delta uk}\sum_{l=0}^{N_e-1}(w_2[l]w_2^*[l]e^{-j\pi\Delta ul})^*.
\end{split}
\end{equation}
The term $\sum_{k=0}^{M_e-1}w_1[k]w_1^*[k]e^{-j\pi\Delta uk}$ is the Fourier transform of $w_1[k]w_1^*[k],$ and it is given by the periodic convolution of $W_1(\Delta u)$ with $W_1^*(-\Delta u)$. Similarly, the term $\sum_{l=0}^{N_e-1}w_2[l]w_2^*[l]e^{-j\pi\Delta ul}$ is the Fourier transform of $w_2[l]w_2^*[l]$, and it is given by the periodic convolution of $W_2(\Delta u)$ with $W_2^*(-\Delta u)$. Thus,  $B$ simplifies to
\begin{equation}
\label{covarianceeq}
B=\dfrac{\sigma^4}{|\norm|^2} W_1(\Delta u)\circledast W_1^*(-\Delta u)W_2^*(\Delta u)\circledast W_2(-\Delta u).
\end{equation}

Subtracting the square of the mean, $\sigma^4$, from the second moment cancels the $A$ term, and the covariance becomes $C(\Delta u)=B.$

\subsection{PPO for Multidimensional Arrays and its Expected Value}
\label{sec:multiproof}

With $\mathbf{k}=\mathbf{m}-\mathbf{n}$, the PPO in Section~\ref{sec:multidim} becomes
\begin{equation}
\begin{split}
&\hat{P}(\mathbf{u}) = \dfrac{1}{\gamma}\sum_{kx=-(N_x-1)}^{M_x-1}\sum_{ky=-(N_y-1)}^{M_y-1}\sum_{kz=-(N_z-1)}^{M_z-1}\\
&e^{-j\pi (\mathbf{u}^T\mathbf{k})}\biggl(w_1[\mathbf{k}]x_1[\mathbf{k}]\biggr)\star \biggl(w_2[-\mathbf{k}]x_2[-\mathbf{k}]\biggr)^*.
\end{split}
\end{equation}
This equation emphasizes that the multivariate PPO is the Fourier transform of the function 
\begin{equation}
\nonumber
\begin{split}
\hat{r}[\mathbf{k}]=&\dfrac{1}{\gamma}\biggl(w_1[\mathbf{k}]x_1[\mathbf{k}]\biggr)\star \biggl(w_2[-\mathbf{k}]x_2[-\mathbf{k}]\biggr)^*.
\end{split}
\end{equation}

The expected value of the multivariate ACF estimate is 
\begin{equation}
\nonumber
\begin{split}
&E\lbrace \hat{r}[\mathbf{k}]\rbrace= E\biggl\lbrace \dfrac{1}{\gamma}(w_1[\mathbf{k}]x_1[\mathbf{k}])\star (w_2[-\mathbf{k}]x_2[-\mathbf{k}])^*\biggr\rbrace\\
=&E\biggl\lbrace \dfrac{1}{\gamma}\sum_{l_x}\sum_{l_y}\sum_{l_z}(w_1[\mathbf{l}]x_1[\mathbf{l}]){(w_2[\mathbf{l}-\mathbf{k}]x_2[\mathbf{l}-\mathbf{k}])^*\biggr\rbrace}\\
\end{split}
\end{equation}
\begin{equation}
\nonumber
\begin{split}
=&\dfrac{r_{xx}[\mathbf{k}]}{\gamma}w_1[\mathbf{k}]\star w_2^*[-\mathbf{k}]=r_{xx}[\mathbf{k}]w_c[\mathbf{k}],
\end{split}
\end{equation}

where $w_c[\mathbf{k}]$ is defined in  \eqref{multiwc}.

\bibliographystyle{IEEEtran}
\bibliography{IEEEabrv,referencesGPfSSE}

\begin{thebibliography}{10}
\providecommand{\url}[1]{#1}
\csname url@samestyle\endcsname
\providecommand{\newblock}{\relax}
\providecommand{\bibinfo}[2]{#2}
\providecommand{\BIBentrySTDinterwordspacing}{\spaceskip=0pt\relax}
\providecommand{\BIBentryALTinterwordstretchfactor}{4}
\providecommand{\BIBentryALTinterwordspacing}{\spaceskip=\fontdimen2\font plus
\BIBentryALTinterwordstretchfactor\fontdimen3\font minus
  \fontdimen4\font\relax}
\providecommand{\BIBforeignlanguage}[2]{{%
\expandafter\ifx\csname l@#1\endcsname\relax
\typeout{** WARNING: IEEEtran.bst: No hyphenation pattern has been}%
\typeout{** loaded for the language `#1'. Using the pattern for}%
\typeout{** the default language instead.}%
\else
\language=\csname l@#1\endcsname
\fi
#2}}
\providecommand{\BIBdecl}{\relax}
\BIBdecl

\bibitem{VanTrees}
H.~V. Trees, \emph{Optimum Array Processing (Detection, Estimation and
  Modulation Theory, Part IV)}.\hskip 1em plus 0.5em minus 0.4em\relax John
  Wiley and Sons, Inc., New York, 2002.

\bibitem{JD}
D.~Johnson and D.~Dudgeon, \emph{Array Signal Processing: Concepts and
  Techniques}.\hskip 1em plus 0.5em minus 0.4em\relax Simon \& Schuster, 1992.

\bibitem{HoctorKassam}
R.~Hoctor and S.~Kassam, ``The unifying role of the coarray in aperture
  synthesis for coherent and incoherent imaging,'' \emph{Proceedings of the
  IEEE}, vol.~78, no.~4, pp. 735--752, April 1990.

\bibitem{berman}
A.~Berman and C.~S. Clay, ``Theory of time averaged product arrays,'' \emph{The
  Journal of the Acoustical Society of America}, vol.~29, no.~7, pp. 805--812,
  1957.

\bibitem{nested1}
P.~Pal and P.~Vaidyanathan, ``Nested arrays: A novel approach to array
  processing with enhanced degrees of freedom,'' \emph{IEEE Transactions on
  Signal Processing}, vol.~58, no.~8, pp. 4167 --4181, {A}ugust 2010.

\bibitem{nested2}
------, ``Nested arrays in two dimensions, part {I}: Geometrical
  considerations,'' \emph{IEEE Transactions on Signal Processing}, vol.~60,
  no.~9, pp. 4694 --4705, {S}eptember 2012.

\bibitem{VandP1}
P.~Vaidyanathan and P.~Pal, ``Sparse sensing with co-prime samplers and
  arrays,'' \emph{IEEE Transactions on Signal Processing}, vol.~59, no.~2, pp.
  573--586, {F}ebruary 2011.

\bibitem{VandP2}
------, ``Theory of sparse coprime sensing in multiple dimensions,'' \emph{IEEE
  Transactions on Signal Processing}, vol.~59, no.~8, pp. 3592--3608, {A}ugust
  2011.

\bibitem{psdestimation}
K.~Adhikari and J.~Buck, ``Spatial spectral estimation with product processing
  of a pair of colinear arrays,'' \emph{IEEE Transactions on Signal
  Processing}, vol.~65, no.~9, pp. 2389--2401, May 2017.

\bibitem{Kay}
S.~Kay, \emph{Modern Spectral Estimation Theory and Application}.\hskip 1em
  plus 0.5em minus 0.4em\relax Prentice Hall, Englewood Cliffs, NJ, 1988.

\bibitem{KBW}
K.~Adhikari, J.~Buck, and K.~Wage, ``Beamforming with extended co-prime sensor
  arrays,'' \emph{2013 IEEE International Conference on Acoustics, Speech and
  Signal Processing (ICASSP)}, pp. 4183--4186, May 2013.

\bibitem{shadings}
\BIBentryALTinterwordspacing
------, ``Extending coprime sensor arrays to achieve the peak side lobe height
  of a full uniform linear array,'' \emph{EURASIP Journal on Advances in Signal
  Processing}, vol. 2014, no.~1, p. 148, {S}ep 2014. [Online]. Available:
  \url{https://doi.org/10.1186/1687-6180-2014-148}
\BIBentrySTDinterwordspacing

\bibitem{MartinoIodice}
G.~{Di Martino} and A.~{Iodice}, ``Coprime synthetic aperture radar (copsar): A
  new acquisition mode for maritime surveillance,'' \emph{IEEE Transactions on
  Geoscience and Remote Sensing}, vol.~53, no.~6, pp. 3110--3123, June 2015.

\bibitem{adhikariJasa1}
\BIBentryALTinterwordspacing
K.~Adhikari, ``Beamforming with semi-coprime arrays,'' \emph{The Journal of the
  Acoustical Society of America}, vol. 145, no.~5, pp. 2841--2850, 2019.
  [Online]. Available: \url{https://doi.org/10.1121/1.5100281}
\BIBentrySTDinterwordspacing

\bibitem{chavali2}
V.~Chavali, K.~Wage, and J.~Buck, ``Multiplicative and min processing of
  experimental passive sonar data from thinned arrays,'' \emph{journal = {The
  Journal of the Acoustical Society of America}}, vol. 144, no.~6, pp.
  3262--3274, December 2018.

\bibitem{icasspdetection}
K.~Adhikari and J.~Buck, ``Gaussian signal detection by coprime sensor
  arrays,'' \emph{2015 IEEE International Conference on Acoustics, Speech and
  Signal Processing (ICASSP)}, pp. 2379--2383, April 2015.

\bibitem{liubuck4}
Y.~{Liu} and J.~R. {Buck}, ``Gaussian source detection and spatial spectral
  estimation using a coprime sensor array with the min processor,'' \emph{IEEE
  Transactions on Signal Processing}, vol.~66, no.~1, pp. 186--199, Jan 2018.

\bibitem{detection2019}
K.~Adhikari and J.~Buck, ``Gaussian signal detection with product arrays,''
  \emph{IEEE Access}, vol.~7, pp. 36\,256--36\,266, 2020.

\bibitem{WageMagazine}
K.~{Wage}, ``When two wrongs make a right: Combining aliased arrays to find
  sound sources,'' \emph{Acoustics Today}, vol.~14, no.~3, pp. 48--56, 2018.

\bibitem{AdhikariAccess1}
K.~{Adhikari} and B.~{Drozdenko}, ``Design and statistical analysis of tapered
  coprime and nested arrays for the min processor,'' \emph{IEEE Access},
  vol.~7, pp. 139\,601--139\,615, 2019.

\bibitem{AccessCoprime1}
W.~{Si}, F.~{Zeng}, C.~{Zhang}, and Z.~{Peng}, ``Improved coprime arrays with
  reduced mutual coupling based on the concept of difference and sum coarray,''
  \emph{IEEE Access}, vol.~7, pp. 66\,251--66\,262, 2019.

\bibitem{asaconf1}
\BIBentryALTinterwordspacing
P.~Johnson, D.~Sartori, T.~Trosclair, J.~Willis, and K.~Adhikari,
  ``Multiplicity of coprime pairs for extension of coprime sensor arrays,''
  \emph{The Journal of the Acoustical Society of America}, vol. 145, no.~3, pp.
  1733--1733, 2019. [Online]. Available:
  \url{https://doi.org/10.1121/1.5101364}
\BIBentrySTDinterwordspacing

\bibitem{AccessCoprime2}
A.~M.~A. {Shaalan} and X.~{Yu}, ``Doa estimation based on the optimized coprime
  array configuration,'' \emph{IEEE Access}, vol.~7, pp. 38\,789--38\,797,
  2019.

\bibitem{asaconf2}
\BIBentryALTinterwordspacing
H.~Elsaadawy, K.~M. Houte, C.~LeBlanc, J.~M. Slezak, and K.~Adhikari, ``Nested
  sensor array extension factors required to match the peak sidelobe height of
  a uniform linear array,'' \emph{The Journal of the Acoustical Society of
  America}, vol. 145, no.~3, pp. 1733--1733, 2019. [Online]. Available:
  \url{https://doi.org/10.1121/1.5101363}
\BIBentrySTDinterwordspacing

\bibitem{AccessCoprime6}
H.~{Xu}, D.~{Wang}, B.~{Ba}, W.~{Cui}, and Y.~{Zhang}, ``Direction-of-arrival
  estimation for both uncorrelated and coherent signals in coprime array,''
  \emph{IEEE Access}, vol.~7, pp. 18\,590--18\,600, 2019.

\bibitem{AccessCoprime8}
H.~{Zhai}, X.~{Zhang}, W.~{Zheng}, and P.~{Gong}, ``Doa estimation of
  noncircular signals for unfolded coprime linear array: Identifiability, dof
  and algorithm (may 2018),'' \emph{IEEE Access}, vol.~6, pp. 29\,382--29\,390,
  2018.

\bibitem{AccessCoprime11}
J.~{Li} and X.~{Zhang}, ``Direction of arrival estimation of quasi-stationary
  signals using unfolded coprime array,'' \emph{IEEE Access}, vol.~5, pp.
  6538--6545, 2017.

\bibitem{AccessCoprime17}
W.~{Si}, Y.~{Wang}, and C.~{Zhang}, ``Three-parallel co-prime polarization
  sensitive array for 2-d doa and polarization estimation via sparse
  representation,'' \emph{IEEE Access}, vol.~7, pp. 15\,404--15\,413, 2019.

\bibitem{AdhikariNAECON}
K.~{Adhikari} and B.~{Drozdenko}, ``Comparison of {M}{U}{S}{I}{C} variants for
  sparse arrays,'' \emph{2019 IEEE National Aerospace and Electronics
  Conference (NAECON)}, pp. 398--405, 2019.

\bibitem{AdhikariAccess2}
------, ``Symmetry-imposed rectangular coprime and nested arrays for direction
  of arrival estimation with multiple signal classification,'' \emph{IEEE
  Access}, vol.~7, pp. 153\,217--153\,229, 2019.

\bibitem{JenkinsWatts}
G.~Jenkins and D.~Watts, \emph{Spectral Analysis and Its Applications}.\hskip
  1em plus 0.5em minus 0.4em\relax Holden-Day, San Francisco, CA, 1968.

\end{thebibliography}

\end{document}